# Quantum Annealing Based Binary Compressive Sensing with Matrix Uncertainty

Ramin Ayanzadeh[1], Seyedahmad Mousavi[2], Milton Halem[1] and Tim Finin[1]

[1] Department of Computer Science and Electrical Engineering, University of Maryland, Baltimore County, Baltimore MD 21250, USA
ayanzadeh@umbc.edu
[2] Department of Mathematics and Statistics, University of Maryland, Baltimore County, Baltimore MD 21250, USA

**Abstract.** Compressive sensing is a novel approach that linearly samples sparse or compressible signals at a rate much below the Nyquist-Shannon sampling rate and outperforms traditional signal processing techniques in acquiring and reconstructing such signals. Compressive sensing with matrix uncertainty is an extension of the standard compressive sensing problem that appears in various applications including but not limited to cognitive radio sensing, calibration of the antenna, and deconvolution. The original problem of compressive sensing is NP-hard so the traditional techniques, such as convex and nonconvex relaxations and greedy algorithms, apply stringent constraints on the measurement matrix to indirectly handle this problem in the realm of classical computing. We propose well-posed approaches for both binary compressive sensing and binary compressive sensing with matrix uncertainty problems that are tractable by quantum annealers. Our approach formulates an Ising model whose ground state represents a sparse solution for the binary compressive sensing problem and then employs an alternating minimization scheme to tackle the binary compressive sensing with matrix uncertainty problem. This setting only requires the solution uniqueness of the considered problem to have a successful recovery process, and therefore the required conditions on the measurement matrix are notably looser. As a proof of concept, we can demonstrate the applicability of the proposed approach on the D-Wave quantum annealers; however, we can adapt our method to employ other modern computing phenomena–like adiabatic quantum computers (in general), CMOS annealers, optical parametric oscillators, and neuromorphic computing.

## 1 Introduction

Digital systems employ analog-to-digital converters (ADC) and digital-to-analog converters (DAC) for acquiring data from the environment and representing the computational results respectively. From a signal processing point of view, the Nyquist–Shannon theorem (also known as sampling theorem) plays a cornerstone role for bridging the gap between the continues-time (analog) and discrete-time (digital) signals. The sampling theorem introduces the rate which is sufficient for perfect reconstruction of the desired signal (Rabiner and Gold 1975).

In real-world applications, remarkable portions of original signal sources are either sparse or compressible (possibly after a change of basis); therefore, traditional signal acquisition methods result in too many samples compared to the actual information contained in the original signal (Foucart and Rauhut 2013). Also, when we perform necessary transforms on digital signals, the majority of transformed coefficients are discarded, and we only retain larger ones for storage and transmission purposes (Rani, Duhok, and Deshmukh 2018). Moreover, in several applications like multiband signals having wide spectral range, the suggested sampling rate by sampling theorem can exceed the specifications of best available analog-to-digital converters (Rani, Duhok, and Deshmukh 2018). A novel tool to resolve these drawbacks is compressive sensing where linear measurements are utilized instead of point measurements.

Compressive sensing (also known as compressed sensing, compressive sampling or sparse sampling) is a recent sensing approach that outperforms the Nyquist–Shannon technique in acquiring and reconstructing sparse and compressible signals (Candès, Romberg, and Tao 2006; Donoho 2006; Candes and Tao 2006). From an application point of view, compressive sensing has demonstrated outstanding performance where: (1) we are restricted by the factor of energy consumption on sensing side (e.g., wireless sensor networks); (2) we are limited to use few sensors (i.e., non-visible wavelengths); (3) sensing is time-consuming (like medical imaging); or (4) measurement/sensing is too expensive (e.g., high-speed ADC). From signal processing attitude, compressive sensing exploits the sparsity of signals through optimization methods to reconstruct the original signal from far fewer samples than the imposed rate by the sampling theorem (Foucart and Rauhut 2013).

Compressive sensing acquires only a small number of random linear measurements of a sparse or compressible signal and subsequently employs $\ell_p$-minimization techniques (generally with $p = 1$) to reconstruct the signal in its original domain. The standard compressive sensing assumes that the measurements come from noiseless sources and we know the linear transform/operator perfectly; however, such assumptions are not valid in the majority of real-world applications. For handling noisy measurements, it is advantageous to apply the idea of penalty methods and extend the measurements to include an additive noise (like white noise). It is often the case that we also do not have complete information about the linear operator due to various reasons –e.g., imperfect calibration, model mismatching, parameter discretization or imperfect signal acquisition hardware. The problem of compressive sensing with matrix uncertainty is able to reconstruct sparse or compressible signals when we only have partial prior information about the measurement matrix (or we only know its approximation) and has widespread applications in scientific and engineering disciplines including but not limited to deconvolution, cognitive radio sensing and calibration of the antenna. From problem-solving attitude, the original problem of compressive sensing and its extensions (i.e., compressive sensing with matrix uncertainty) are intractable in the realm of classical computing (Muthukrishnan 2005). Therefore, we impose additional severe constraints on the measurement matrix to attack the NP-hard compressive sensing problem through convex optimization so that fast interior methods can tackle them efficiently.

Quantum annealers are adiabatic quantum computers that tackle discrete optimization problems via minimizing an Ising model whose ground state represents the solution for the problem of interest (Kadowaki and Nishimori 1998). Considering trends in commercializing quantum annealing machines by D-Wave Systems, achieving global optimum for discrete optimization problems seems tangible. After introducing the first programmable quantum machine by D-Wave Systems (Johnson et al. 2011) with 128 quantum bits (qubits), the number of available qubits in D-Wave's quantum annealer has been increased significantly. The current D-Wave quantum annealer (introduced in fall 2017) has more than 2,000 qubits, and the next generation of it will include more than 5,000 qubits. D-Wave quantum annealer can minimize any energy function in Quadratic Unconstrained Binary Optimization (QUBO) form or its equivalent Ising model (O'Malley et al. 2017). Although experimental results have revealed that the D-Wave quantum annealer can find very high quality global optimum solutions in a fraction of second, it is challenging to not only represent the problems as an Ising model but also map the objective functions to a sparse coefficient matrix (chimera topology).

In this study, we reduce the standard problem of binary compressive sensing to an Ising model whose ground state represents a favorable binary sparse signal. In other words, we propose a well-posed binary compressive sensing that employs quantum annealers to tackle the $\ell_0$-minimization problem directly. Hence, our approach only requires the solution uniqueness of this problem for which the required conditions on the measurement matrix are much looser. For solving the binary compressive sensing with matrix uncertainty problem, we leverage our proposed Ising model via an alternating minimization scheme to predict both a desired sparse signal and an uncertainty parameter vector simultaneously. As a proof of concept, we can demonstrate that we can map the resulting abstract Ising models on the D-Wave quantum annealer hardware; however, we can adapt it to employ other modern computing phenomena –including but not limited to CMOS annealers, neuromorphic computing and optical parametric oscillators.

## 2  Problem Definition and Related Works

The original problem of compressive sensing aims to recover a sparse signal $x \in \mathbb{R}^N$ from a given measurement vector $y \in \mathbb{R}^m$ such that $y = Ax$ and the measurement matrix $A \in \mathbb{R}^{m \times N}$ with $m \ll N$ (this matrix is also known as coding matrix, design matrix or linear transform/operator). This problem can have infinite solutions, and compressive sensing guarantees the uniqueness of sparse solutions under different conditions. Let $\|x\|_0$ stand for the sparsity level of $x$, i.e., the number of nonzero entries of $x$, the ultimate goal of compressive sensing can be formulated as follows:

$$\min_{x \in \mathbb{R}^N} \|x\|_0 \quad \text{subject to} \quad Ax = y. \tag{1}$$

This problem is NP-hard (Muthukrishnan 2005) due to its combinatorial nature in selecting the best support set and consequently, it is not tractable and must be handled indirectly in the realm of classical computing. Because $\ell_p$-norm approximates $\ell_0$-norm as nonnegative $p$ goes down to zero, the standard approach to tackle this problem is exploiting the following $\ell_p$-norm problem:

$$\min_{x \in \mathbb{R}^N} \| x \|_p \quad \text{subject to} \quad Ax = y; \qquad (2)$$

and then investigating that under what conditions and for which $p$'s problems (1) and (2) appoint an identical solution. Various results certify this possibility using tools such as restricted isometry constants or null and range space properties. For $p > 1$ the unique solution of this problem is not sparse and each of its components is nonzero except in pathological situations (Shen and Mousavi 2018). The case of $p = 1$ leads to a linear program; so it does not necessarily attain a unique solution, although there are verifiable necessary and sufficient conditions under which a feasible point is guaranteed to be the unique solution (Mousavi and Shen 2017). But this convex program and the original nonconvex $\ell_0$-norm problem appoint an identical and unique sparse solution under appropriate circumstances (Foucart and Rauhut 2013). For instance, the mutual coherence of the measurement matrix $A$ is defined below:

$$\mu(A) = \max_{i \neq j} \frac{|\langle A_i, A_j \rangle|}{\| A_i \|_2 \cdot \| A_j \|_2}$$

This simply seeks the largest correlation between two different columns and provides a sufficient condition for the equivalence of the $\ell_0$ and $\ell_1$-norm problems. More specifically, $0 \neq x \in \mathbb{R}^N$ is an identical and unique solution for both problems (1) and (2) when $\mu(A) < 0.5(1 + \mu(A)^{-1})$ (Elad and Bruckstein 2002). This property provides an improper upper bound because it is practical only when the original signal is highly sparse.

The measurement matrix $A$ has the restricted isometry property (RIP) of order $s$ if there exists $\delta_s \in (0,1)$ that satisfies the following:

$$(1 - \delta_s) \| x \|_2^2 \leq \| Ax \|_2^2 \leq (1 + \delta_s) \| x \|_2^2 \qquad \forall x; \| x \|_0 \leq s.$$

The $\ell_0$-norm and $\ell_1$-norm problems appoint the same $s$-sparse (at most $s$ nonzero entries) solution if $\delta_{2s} < 0.4931$ (Mo and Li 2011). The RIP demands each column submatrix $A_S$ with $card(S) \leq s$ to behave like an identity matrix, precisely, to have singular values in $[1 - \delta_s, 1 + \delta_s]$. Since this property involves all the $s$-tuples of columns of $A$, it is more rigorous than the mutual coherence (in fact, $\delta_2 = \mu$) so that it derives to better upper bounds on the sparsity level of a vector to be recovered. From sparsity point of view, the RIP enables the compressive sensing to recover remarkably less sparse signals –compare to the mutual coherence property (Mo and Li 2011).

Although the choice of $p=1$ is the most interesting case as $\ell_1$ is the closest convex norm to $\ell_0$ (Ramirez, Kreinovich, and Argaez 2013) and convex optimization is extremely well-nourished, the shape of a unit ball associated with $\ell_p$-norm for $0<p<1$ motivates researchers to explore this case. For certain values of such $p$, we can obtain not only more robust and stable theoretical guarantees but also much less restrictive conditions than $\ell_1$-norm recovery (Saab, Chartrand, and Yilmaz 2008; Chartrand 2009). For instance, a sufficient condition for recovering an $s$-sparse vector in the noiseless case via $\ell_{0.5}$ minimization is $\delta_{3s}+27\delta_{4s}<26$, which is remarkably less restrictive than the analogous result for $\ell_1$-norm recovery that requires $\delta_{2s}+2\delta_{3s}<1$ (Saab, Chartrand, and Yilmaz 2008).

Compressive sensing relies on the sparsity of signals and linearly samples them at a rate much below the Nyquist sampling rate; however, majority of signals in real-world applications are not precisely sparse in their original domain. Since the majority of signals have approximately sparse representation in some transformed domain (e.g., wavelet domain for natural images, Fourier domain for speech signals, Radon domain for medical images, etc.), we extend $\ell_p$-norm recovery analysis to the case of compressible vectors that are practically more valuable (Rani, Duhok, and Deshmukh 2018). Roughly speaking, a vector $x$ is compressible or nearly $s$-sparse in $\ell_p$-norm if $\|x-x_S\|_p$ is small enough such that $S$ is an index set containing its $s$ largest absolute components. Besides, for handling noisy measurements, it is advantageous to apply the idea of penalty methods and attack this NP-hard unconstrained problem:

$$\min_{x\in\mathbb{R}^N} \|y-Ax\|_2^2 + \lambda \|x\|_0. \qquad (3)$$

There is a trade-off situation between feasibility and sparsity in the nature of this problem that we control it by the penalty parameter $\lambda$ (Figueiredo, Nowak, and Wright 2007).

Majority of results for recovering $s$-sparse signals in compressive sensing literature are based on the RIP constants; specifically in the form of $\delta_s$ or $\delta_{ks} \leq \delta$ for natural number $k>0$ and $\delta \in (0,1)$. The probabilistic nature in such results for the standard problem of compressive sensing, as an important special case of compressive sensing with matrix uncertainty problem, is not favorable because such conditions are not indeed verifiable, due to its computational complexity (Tillmann and Pfetsch 2014). Hence, there is an interest in constructing deterministic matrices with small RIP constants, but the number of measurements $m$ scales quadratically in sparsity level $s$ in the ongoing results. A breakthrough in this field took place when random matrices demonstrated their capability to satisfy such appealing inequalities with a high probability. As an illustration, for an $m \times N$ random matrix, where each entry is independently drawn according to Gaussian or Bernoulli distributions, we have $\delta_s \leq \delta^*$ for $m \geq C\delta^{-2} s \ln(\frac{eN}{s})$ in which $C>0$ does not depend on $s, m$ and $N$ (Baraniuk et al.

62008). It is worth noting that various applications, however, impose a measurement matrix that is not necessarily random. As a result, the recovery process is not guaranteed even with a high probability.

In standard compressive sensing, a major assumption is to have a perfect measurement matrix, although this is not appropriate for many vital practical circumstances such as the erroneous measurement matrices that emerge in parameter discretization, telecommunication, source separation, and model mismatch (Herman and Strohmer 2009; Fannjiang, Strohmer, and Yan 2010). In other words, a realistic model not only accounts for the measurement error $y$ but also handles numerical errors in the measurement matrix $A$. A well-documented approach to model this issue is considering an additive perturbation for the measurement matrix and minimizing the magnitude of this unknown uncertainty matrix along with the measurement error. There is an extensive literature to investigate the impact of uncertainty assumption on the existing efficient algorithms in compressive sensing including Basis Pursuit and Compressive Sampling Matching Pursuit (Zhu, Leus, and Giannakis 2011; Herman and Strohmer 2010; Chi et al. 2011). Compressive sensing with matrix uncertainty, associated to the Bayesian setting with prior informative, is a more recent approach that aims to reconstruct a sparse or compressible signal $x \in \mathbb{R}^N$ and finding an uncertainty parameter vector $d \in \mathbb{R}^r$ from a noisy measurement vector $y \in \mathbb{R}^m$ such that $y = A(d)x + e$ where $A(d) = A_0 + \sum_{i=1}^{r} d_i A_i$ and $A_i \in \mathbb{R}^{m \times N}; \forall i,$ and the unknown noise vector $e \sim \mathcal{N}(0, I/\gamma)$ (Parker, Cevher, and Schniter 2011; Meng and Zhu 2018).

## 3   Quantum Annealing

Quantum computing is a novel approach for tackling computationally intensive problems that are intractable for classical computers. In conventional computers, data are encoded as a sequence of bits that take their value from $\{0,1\}$ so a register with $n$ bit at time $t$ can be in only one of the $2^n$ possible states. Solving a problem on classical computers requires a sequence of Boolean operations that can become intractable when the state space grows exponentially. On the other side, a quantum bit (qubit) in a quantum computer is a two-level quantum system that can take its value from $|0\rangle, |1\rangle$ (analogous to classical bits) as well as any superposition of the basis $|0\rangle$ and $|1\rangle$. In general, the state of a qubit is represented as:

$$|\psi\rangle = \alpha |0\rangle + \beta |1\rangle$$

where $\alpha$ and $\beta$ are complex coefficients of the system. Performing the measurement on qubits collapses their quantum state to $0$ or $1$ with the probability of $|\alpha|^2$ and $|\beta|^2$ respectively. After introducing the concept of quantum computing by Richard Feynman in 1959, several models have been proposed for the realization of quantum information processing. Although these models are theoretically equivalent, their underlying concepts, as well as realization requirements, are significantly different from each other. From an implementation perspective, quantum computers are categorized as universal and adiabatic devices.

In universal quantum computing, quantum bits (qubits) are based on well-defined quantum states, and quantum algorithms are executed based on the quantum equivalent of the logic gates. Having qubits in superposition and taking the advantages of their entanglement, we can apply single-qubit quantum operations (e.g., Hadamard, Phase, rotation and $\pi/8$) and two-qubit quantum operations (e.g., Toffoli and CNOT) to achieve the quantum supremacy. Circuit models in quantum computing are closely analogous to classical (transistor-based) computers in which, the sequence of single-qubit and two-qubit operations (quantum gates) are applied on quantum registers to perform the algorithmic quantum information processing tasks. Unlike classical gates that are not reversible (except the NOT and Toffoli gates), quantum gates are reversible (Lamonica 2002a). Since the measurement operation is not reversible, we consider it as a non-quantum operator.

Since circuit models resemble the quantum information processing through a sequence of quantum operations, the circuit must have enough coherence time for running the quantum algorithms (Lomonaco 2002b). In some sense, it is difficult not only to design problem specific quantum algorithms but also fabricate large-scale quantum circuits for real-world applications. Although progress in developing quantum computers look promising, universal quantum computing is less likely to be able to tackle real-world problems in the short future. As an illustration, factoring a 2048 bits integer using Shor's algorithm requires about 4,100 error-free qubits, and due to massive error correction computations, we will need more than 40 million qubits to jeopardize the security of current RSA based public-key cryptography systems. It is worth noting that the most recently developed quantum circuits are limited to only 50 qubits (by IBM) and 72 qubits (by Google). The Rigetti has announced to deliver a 128-qubit system in 2019.

Adiabatic quantum computers rely on the "Quantum Adiabatic Evolution" for processing the quantum information. According to the "Adiabatic Theorem," adiabatic quantum computers are polynomially equivalent to the circuit models. Unlike previous models in quantum information processing, qubits of adiabatic quantum computers do not perform discrete operations. Indeed, adiabatic quantum computers receive a Hamiltonian (also called energy function) as input whose ground state represents the solution for the problem that we are trying to solve. Afterward, qubits are adiabatically evolved from some initial unknown state to a final state that minimizes the energy function (Farhi et al. 2000). Quantum annealers are the type of adiabatic quantum computers that are mainly used for discrete optimization problems (Kadowaki and Nishimori 1998). Quantum annealers do not have enough coherence time to run quantum algorithms but they are significantly simpler than digital quantum computers, so they are remarkably easier to scale.

The D-Wave quantum annealer is an Ising processing unit (IPU) which minimizes any energy function in Quadratic Unconstrained Binary Optimization (QUBO) form or its equivalent Ising model (Johnson et al. 2011). More precisely, D-Wave quantum annealer receives the coefficients of an Ising model as input (here $h$ and $J$) and returns the vector $z$ (where $z_i \in \{-1,+1\}$) that minimizes the energy function $E$ as below (O'Malley et al. 2017):

$$E(z) = \sum_i h_i z_i + \sum_{i<j} J_{i,j} z_i z_j. \tag{4}$$

The D-Wave quantum annealer architecture has some limitations that restrict the process of mapping problems into an executable quantum machine instruction. Although the current generation of the D-Wave quantum annealer includes more than 2,000 qubits, it only has about 6,000 couplings; meaning, the connectivity architecture is sparse. Figure 1 illustrates the topology of this connectivity (known as Chimera graph). In practice, we can entangle multiple qubits to form a virtual qubit (also called chaining) with higher connectivity. It is worth highlighting that the next generation of the D-Wave quantum annealer will include more than 5,000 qubits and 40,000 couplings.

Programming the current generations of the D-Wave quantum annealer also requires additional constraints on the coefficients of the input Ising model. More narrowly, $h_i \in [-2,2]$ and $J_{i,j} \in [-1,+1]$, and precision of these coefficients is limited to only 4-5 bits. Recent studies have proposed practical heuristics for addressing this issue (Dorband 2018b). Moreover, owing to technology barriers, the D-Wave quantum annealer cannot guarantee to find the global minimum of the given Hamiltonian. On this basis, several post-processing heuristics tackle this issue to enhance the performance of the D-Wave quantum annealer (Dorband 2018a).

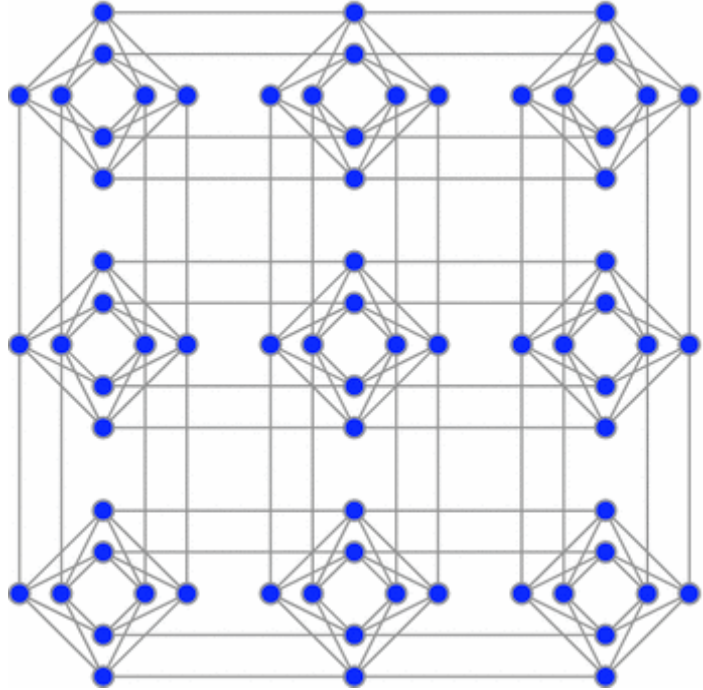

Fig. 1. Chimera Topology in D-Wave Quantum Annealers

# 4 Quantum Annealing Based Binary Compressive Sensing

Sparse recovery tries to infer high-dimensional sparse or compressible objects including vectors, matrices and high-order tensors from very limited observations. From problem-solving point of view, sparse recovery has to deal with not only $\ell_0$-norm problem but also large-scale matrices that generally have remarkably more columns. In other words, we must deal with combinatorial problems that have large dimensions. The standard technique to tackle the $\ell_0$ problem is to replace it by $\ell_1$ and solve the new (possibly) convex problem. In compressive sensing, we are interested in matrices for which these problems attain an identical solution. These conditions are mostly NP-hard to verify, and also may not hold in some applications. Further, the size of the new approximation problem is prohibitive even if efficient interior-point methods are applied.

Quantum annealing based binary compressive sensing aims to employ adiabatic quantum computers (here quantum annealers) to tackle the $\ell_0$-minimization problem directly (3) and reconstruct a binary sparse or compressible signal from a small number of linear measurements. The D-Wave quantum annealer receives the coefficients of an Ising model as input (here $h$ and $J$) and returns the vector $z$ ($z_i \in \{-1,+1\}$) that minimizes the corresponding Ising model (or energy function) in equation (4).

To find the associated coefficients in equation (4), we first formulate the problem of binary compressive sensing as a quadratic unconstrained binary optimization (QUBO) and then apply a linear transform to form an Ising model whose ground state represents a sufficiently sparse solution. Here, our ultimate goal is to employ the D-Wave quantum annealer for minimizing the objective function $\| y - Ax \|_2^2 + \lambda \| x \|_0$, with $x_i \in \{0,1\}$. Hence, we need to rewrite $f(x) := -2y^T Ax + x^T A^T Ax + \lambda \| x \|_0$ in the QUBO form, that is, $f(x) = \sum_i h_i x_i + \sum_{i,j:i<j} J_{ij} x_i x_j$. We start with the simple quadratic function of $x^T Bx$ for a square matrix $B$. First, note that

$$x^T Bx = \sum_i x_i (Bx)_i = \sum_i x_i \left( \sum_j B_{ij} x_j \right) = \sum_{i,j} B_{ij} x_i x_j.$$

Thus, if $B = B^T$, the above reduces to

$$\sum_{i,j} B_{ij} x_i x_j = \sum_i B_{ii} x_i^2 + \sum_{i,j:i<j} B_{ij} x_i x_j + \sum_{i,j:i>j} B_{ji} x_i x_j = \sum_i B_{ii} x_i^2 + \sum_{i,j:i<j} 2 B_{ij} x_i x_j.$$ Since $x_i \in \{0,1\}$, we have $x^T Bx = \sum_i B_{ii} x_i + \sum_{i,j:i<j} 2 B_{ij} x_i x_j$. Thus, we conclude that

$$f(x) = -2y^T Ax + x^T A^T Ax + \lambda \sum_i x_i$$

$$= \sum_i \left( \sum_l -2y_l A_{li} + \lambda \right) x_i + x^T A^T Ax$$

$$= \sum_i \left( \sum_l -2y_l A_{li} + \lambda \right) x_i + \sum_i B_{ii} x_i + \sum_{i,j:i<j} 2 B_{ij} x_i x_j \quad (where \quad B = A^T A)$$

$$= \sum_i \left( \sum_l -2y_l A_{li} + \lambda \right) x_i + \sum_i \left( \sum_l A_{li} A_{li} \right) x_i + \sum_{i,j:i<j} \left( 2 \sum_t A_{ti} A_{tj} \right) x_i x_j$$

$$= \sum_i \left( \sum_l -2y_l A_{li} + \lambda + \sum_l A_{li} A_{li} \right) x_i + \sum_{i,j:i<j} \left( 2 \sum_t A_{ti} A_{tj} \right) x_i x_j$$

$$= \sum_i \left( \lambda + \sum_l A_{li} [-2y_l + A_{li}] \right) x_i + \sum_{i,j:i<j} \left( 2 \sum_t A_{ti} A_{tj} \right) x_i x_j.$$

The above implies that

$$\| y - Ax \|_2^2 + \lambda \| x \|_0 - \| y \|_2^2 = f(x) = \sum_i h_i x_i + \sum_{i,j:i<j} J_{ij} x_i x_j;$$

for the following coefficients:

$$h_i = \lambda + \sum_l A_{li} [-2y_l + A_{li}]; \tag{5}$$

$$J_{ij} = 2 \sum_t A_{ti} A_{tj}. \tag{6}$$

In order to map the resulting QUBO formulation of the binary compressive sensing to an Ising model, we can apply a linear transformation as:

$$z_i = 2x_i - 1. \tag{7}$$

The coefficients of the resulting Ising models do not necessarily fall into feasible intervals to form an executable quantum machine instruction on the D-Wave quantum annealer hardware. To handle these infeasible Ising models (based on specific hardware), we can normalize the coefficients simultaneously to avoid this drawback. As an illustration, we can divide the original problem by a positive number that trivially does not change the solution.

## 5  Quantum Annealing Based Binary Compressive Sensing with Matrix Uncertainty

The objective in binary compressive sensing with matrix uncertainty is reconstructing a sparse binary signal $x \in \{0,1\}^N$ and finding an uncertainty parameter vector $d \in \mathbb{R}^r$ from a noisy measurement vector $y \in \mathbb{R}^m$ such that $y = A(d)x + e$ where $A(d) = A_0 + \sum_{i=1}^{r} d_i A_i$ and $A_i \in \mathbb{R}^{m \times N}; \forall i,$ and the unknown noise vector $e \sim \mathcal{N}(0, I/\gamma)$.

We can represent this problem as solving the following NP-hard optimization problem:

$$\min_{x,d} \ \| (A_0 + \sum_{i=1}^{r} d_i A_i)x - y \|_2^2 + \frac{\| d \|_2^2}{\gamma} + \lambda \| x \|_0. \tag{8}$$

This problem entirely inherits favorable properties desired in the compressive sensing with matrix uncertainty by penalizing large sparsity and handling noise in the measurement matrix and vector. In other words, for a given noisy measurement vector $y$, we can adjust the penalty parameter $\lambda$ to control the sparsity level of the resulting solution. The larger is this parameter; the smaller sparsity level is desired so that we can adapt it for specific applications. However, this problem is NP-hard and thus it is not tractable in the realm of classical computing. Here, we leverage the results from the previous section through an alternating minimization scheme to propose a well-posed approach for solving the problem of binary compressive sensing with matrix uncertainty that is tractable by quantum annealers.

Similar to the standard binary compressive sensing, we start with the QUBO representation of the problem and then apply the linear transformation (7), to form the associated Ising model and compile it to an executable quantum machine instruction. To see this, let $d$ be fixed and define $A := A_0 + \sum_{i=1}^{r} d_i A_i$, then the first task is finding the QUBO form of:

$$f_d(x) := \| \left(A_0 + \sum_{i=1}^{r} d_i A_i\right)x - y \|_2^2 + \lambda \| x \|_0 = \| Ax - y \|_2^2 + \lambda \| x \|_0.$$

Since $x_i \in \{0,1\}$, for a fixed uncertainty parameter $d$, we can use equation (5) and equation (6) to rewrite the problem (8) as $f_d(x) - \| y \|_2^2 = \sum_{i=1}^{N} h_i x_i + \sum_{i,j:i<j} J_{ij} x_i x_j$. Next, we assume that $x$ is fixed so that the objective function which needs to be minimized is the following:

$$g_x(d) := \| \left(A_0 + \sum_{i=1}^{r} d_i A_i\right)x - y \|_2^2 + \frac{\| d \|_2^2}{\gamma}.$$

By letting $G := [A_1 x \ A_2 x \ldots A_r x]$ and $c := y - A_0 x$, we have $g_x(d) = \| Gd - c \|_2^2 + \| d \|_2^2 / \gamma$ so that the closed form solution of $\min_d g_x(d)$ can be written as

$$d^* = (G^T G + I/\gamma)^{-1} Gc. \tag{9}$$

Algorithm 1 incorporates the above information and illustrates our proposed well-posed solution for tackling the problem of binary compressive sensing with matrix uncertainty. First, we start our algorithm with initializing the elements of the uncertainty parameter $d$ as zero vector and then use an iterative process in which we alternatively update the vectors $x$ and $d$. Once the vector $d$ is fixed, we use the equations (5) and (6) to rewrite the resulting problem as a QUBO form and then employ the D-Wave quantum annealer to find the ground state of the resulting Ising model. After updating the vector $x$, we assume that it is fixed and use the equation (9) to find the new uncertainty parameter $d$. We repeat the iterations until a termination criterion is met.

**Algorithm 1. Alternating minimization scheme for quantum annealing based binary compressive sensing with matrix uncertainty**

**Input:** $A_0, \ldots, A_r, y,$ and $\epsilon > 0$

**Initialization:** $\lambda > 0,$ and $d = 0$

**Step 1:** Update

$\quad x \leftarrow \arg\min_{x \in \{0,1\}^N} f_d(x)$

$\quad d \leftarrow d^*$

**Step 2:** Stop if $\| (A_0 + \sum_{i=1}^{r} d_i A_i)x - y \|_2 \leq \epsilon$

**Step 3:** Go to step 1

**Output:** sparse vector $x$ and uncertainty parameter $d$

# 6 Discussion

Compressive sensing is a recent approach to recover sparse or compressible signals at a rate significantly below the Nyquist-Shannon sampling rate, which outperforms traditional signal processing techniques. In real-world applications, it is often the case that we have to manage not only noisy measurements but also the imperfect linear operator. Compressive sensing with matrix uncertainty is a general form of the standard compressive sensing problem that meets these goals. The original problem of compressive sensing and its general form (compressive sensing with matrix uncertainty) are NP-hard and we cannot directly solve them in the realm of classical computing. Therefore, it is critical to impose additional severe constraints on the measurement matrices to reformulate them as convex or nonconvex optimization problems.

We propose well-posed approaches for both binary compressive sensing and binary compressive sensing with matrix uncertainty problems that are tractable by quantum annealers. We first directly map the original problem of binary compressive sensing to a quadratic unconstrained binary optimization (QUBO) form and then apply a linear transformation to form an Ising model whose ground state represents a sparse solution for an associated problem of interest. Afterward, we leverage the resulting Ising model via an alternating minimization scheme to propose a well-posed approach to the problem of binary compressive sensing with matrix uncertainty. But the only requirement in our approach is the solution uniqueness of the associated problems which are notably looser.

As a proof of concept, we can demonstrate the applicability of the proposed approach on the D-Wave quantum annealers; however, we can adapt our method to employ other modern computing phenomena–like adiabatic quantum computers (in general), CMOS annealers, optical parametric oscillators and neuromorphic computing. Our method reduces the original problem of binary compressive sensing to an abstract Ising model in which, all qubits are fully connected. The D-Wave quantum annealer, however, has sparse connectivity (Chimera topology). Thus, we need to compile the resulting Ising models to executable quantum machine instructions on the D-Wave quantum annealer. To this end, we can entangle the physical qubits with lower connectivity to form a set of fully connected virtual qubits. It is worth noting that chaining the qubits remarkably ebbs the capacity of the D-Wave quantum annealer. The current generation of the D-Wave quantum annealer with more than 2,000 qubits is equivalent to a fully connected quantum annealer with about 64 qubits. Note that the next generation of the D-Wave quantum annealer will include more than 5,000 qubits and the new Chimera topology will have up to 15 connectivity per qubit (compare to 6 in the current architecture).

## Acknowledgments

This research has been supported by NASA grant (#NNH16ZDA001N-AIST 16-0091), NIH-NIGMS Initiative for Maximizing Student Development Grant (2 R25-GM55036), and the Google Lime scholarship. We would like to thank the D-Wave Systems management team for access to the 2000Q quantum computer at Burnaby, Canada. We also would like to thank the NSF funded Center for Hybrid Multicore Productivity Research for access support to IBM Minsky cluster.